\documentclass[conference]{IEEEtran}
\IEEEoverridecommandlockouts
\usepackage{cite}
\usepackage{amsmath,amssymb,amsfonts}
\usepackage{algorithmic}
\usepackage{graphicx}
\usepackage{textcomp}
\usepackage{xcolor}
\usepackage{multirow}
\usepackage{makecell}
\usepackage{orcidlink}
\def\BibTeX{{\rm B\kern-.05em{\sc i\kern-.025em b}\kern-.08em
    T\kern-.1667em\lower.7ex\hbox{E}\kern-.125emX}}
\begin{document}
\newcommand{\RNum}[1]{\uppercase\expandafter{\romannumeral #1\relax}}

\title{A Global Data-Driven Model for The Hippocampus and Nucleus Accumbens of Rat From The Local Field Potential Recordings (LFP)\\
}

\author{
  \IEEEauthorblockN{
    Maedeh Sadeghi\IEEEauthorrefmark{1}\orcidlink{0000-0002-2096-6369},
    Mahdi Aliyari Shoorehdeli \IEEEauthorrefmark{1},
    Shole jamali\IEEEauthorrefmark{2},
    Abbas Haghparast\IEEEauthorrefmark{2}
  }
  \IEEEauthorblockA{\IEEEauthorrefmark{1} Fault Detection and Identification (FDI) Laboratory, Faculty of Electrical Engineering, \\K. N. Toosi University of Technology, Tehran, Iran}
  \IEEEauthorblockA{\IEEEauthorrefmark{2} Neuroscience Research Center, School of Medicine, Shahid Beheshti University of Medical Sciences,\\ Tehran, Iran}
}

\maketitle
\begin{abstract}
In brain neural networks, Local Field Potential (LFP) signals represent the dynamic flow of information. Analyzing LFP clinical data plays a critical role in improving our understanding of brain mechanisms. One way to enhance our understanding of these mechanisms is to identify a global model to predict brain signals in different situations. This paper identifies a global data-driven based on LFP recordings of the Nucleus Accumbens and Hippocampus regions in freely moving rats. The LFP is recorded from each rat in two different situations: before and after the process of getting a reward which can be either a drug (Morphine) or natural food (like popcorn or biscuit). A comparison of five machine learning methods including Long Short Term Memory (LSTM), Echo State Network (ESN), Deep Echo State Network (DeepESN), Radial Basis Function (RBF), and Local Linear Model Tree (LLM) is conducted to develop this model. LoLiMoT was chosen with the best performance among all methods. This model can predict the future states of these regions with one pre-trained model. Identifying this model showed that Morphine and natural rewards do not change the dynamic features of neurons in these regions.
\end{abstract}


\section{Introduction}
\subsection{Motivation}

Human activities, nature, world affairs, and even the universe can be explained using time series. A time series is a sequence of observations taken at successive intervals. One way to enhance our understanding of time series is to predict their future states. Time series prediction, the process of predicting future states of time series based on past values, is important because it allows us to understand potential futures and explain the past by considering both the present and the future.

Machine learning methods are increasingly used in medical practice to analyze clinical data to uncover patterns and improve understanding \cite{b15}. There are many types of time series in the medical field that can be predicted to give us more information about future activities of specific body parts. Predicting these time series aids researchers in diagnosing possible diseases by understanding the future state of a person's brain signal. One of these signals is Local Field Potentials (LFPs). This signal represents the dynamic flow of information in brain neural networks. Studies have shown that LFP signals are closely related to the individual neurons' activity in the brain \cite{b18}. In recording LFP signals, an electrode is implanted in the brain, and the voltage caused by surrounding neurons is measured.

\subsection{Related Work}
It is a popular research topic to predict time series, including LFP data and other types of brain data. Time series prediction intends to learn more about real cases through the acquired data. Some different methods have been used to model and predict brain signals. Auto-Regressive (AR) model, which is linear time-invariant, is used to model the brain activity of patients with dystonia disease \cite{b13} and epilepsy disease \cite{b28}. In \cite{b6}, the LFPs of three epileptic patients are predicted using a Long Short-Term Memory (LSTM). The trained network on one patient's LFP cannot be used to predict the LFPs of other patients. Reference \cite{b16} models LFP signals with LSTM and \cite{b17} uses a Neural Network based model to detect artifacts of raw LFPs. The human Mirror Neuron System (MNS) is also modeled using a data-driven network model in \cite{b27} with recorded LFP of psychiatric patients. Reference \cite{b14} presents a Dynamic Bayesian Network(DBN) for modeling four different time slices of LFP data collected from rats induced to epilepsy. Various computational models are proposed based on Artificial Neural Networks \cite{b19} like Single Multiplicative Neuron (SMN) \cite{b20}-\cite{b22}, Deep Recurrent Neural Networks (RNN) \cite{b23}, Elman RNN \cite{b24}, LSTM \cite{b25}, and Adaptive Neuro-Fuzzy Inference System (ANFIS) \cite{b26} to predict brain activities of different patients. However, none of them find a global model to describe the dynamical features of a specific region and predict all LFP signals of different cases by a pre-trained model.

\subsection{Contribution}
In this study, our approach is to introduce Local Linear Model Tree (LoLiMoT) network as a global data-driven model for two brain regions of free-moving rats. For this purpose, five methods based on machine learning (LSTM, Echo State Network (ESN), Deep Echo State Network (Deep ESN), Locally Linear model (LLM), and Radial Basis Function (RBF)) are implemented. The performance of LLM is compared to the other networks' performances. To show the globality of this model, the results of signal prediction of 18 rats, which are in different health conditions, are compared. However, to our best knowledge, the LLM method has never been used to predict LFPs or any other types of brain signals such as EEG.

\subsection{Outline}
The rest of the paper is organized as follows: first, recorded data are described in \autoref{data}. \autoref{method} introduces the proposed methodology in detail. In this section, five methods are introduced. The experimental results of these methods are compared in \autoref{result}. And in the last \autoref{con}, the conclusion is presented.

\section{Material and data recording}
\label{data}

The studies showed that Nucleus Accumbens and Hippocampus are essential areas of the brain's reward system \cite{b29}. Thus, analyzing signals of these regions has become a popular research topic to gain a deeper understanding of the reward system. In this study, 19 adult male Wistar weighing between 220g and 270g were kept under a 12h cycle of dark/light with access to water freely. Rats were free to move in two equal-sized chambers but with different wall patterns and floor texture, which was isolated by a guillotine door leading into the third chamber (null chamber). 
The process of recording has three phases. The first phase is known as Pre-test. One day before the conditioning phase, LFP signals were recorded from the Nucleus Accumbens and Hippocampus while rats were free in chambers. Their behavior was monitored with a 3CCD camera. Monitoring was used to check the Conditioning Place Preference (CPP) apparatus.
The second phase is named the conditioning phase. It is a certain period that the animal gets the reward in a specific chamber. Rats were divided into three groups depending on their reward in this phase. Group \RNum{1}, comprising six rats, was rewarded with Morphine. Six rats in group \RNum{2} were rewarded with natural rewards (palatable food like popcorn and biscuits). Seven rats were rewarded with saline in the third group (control group). 
The last phase, called Post-test, is one day after the last day of the conditioning phase. At this time, rats were free again, and LFPs were recorded. Both recordings were obtained for 10 min at a sample rate of 1kHz. The gained LFP signals were amplified 1000 times and filtered by bandpass (0-300 Hz). 

Ethics Committee of Shahid Beheshti University of Medical Sciences (IR.SBMU.SM.REC.1395.373), Tehran, Iran, approved all procedures of experiments, and all protocols followed international standards for the Care and Use of Laboratory Animals (NIH publication No. 80-23, revised in 1996).

\section{Methodology}
\label{method}
\subsection{Long Short Term Memory (LSTM)}
LSTM is a sort of recurrent neural network that does not suffer from other RNN problems such as exploding gradient and vanishing gradient \cite{b11}. The structure of a unit of LSTM is illustrated in \autoref{LSTM}. It has three gates ($g_I(t)$, $g_o(t)$, and $g_F(t)$) and a memory cell $C(t)$. The input gate measures the importance of the network input's new information. Forget gate decides whether the network keeps the information from previous time steps. The memory cell ensures that the gradient can pass through many time steps. The output gate controls the information that passes from the cell to the hidden state. Equations of one step of an LSTM unit are given as follows:

\begin{figure}[htbp]
\centerline{\includegraphics[scale=0.4]{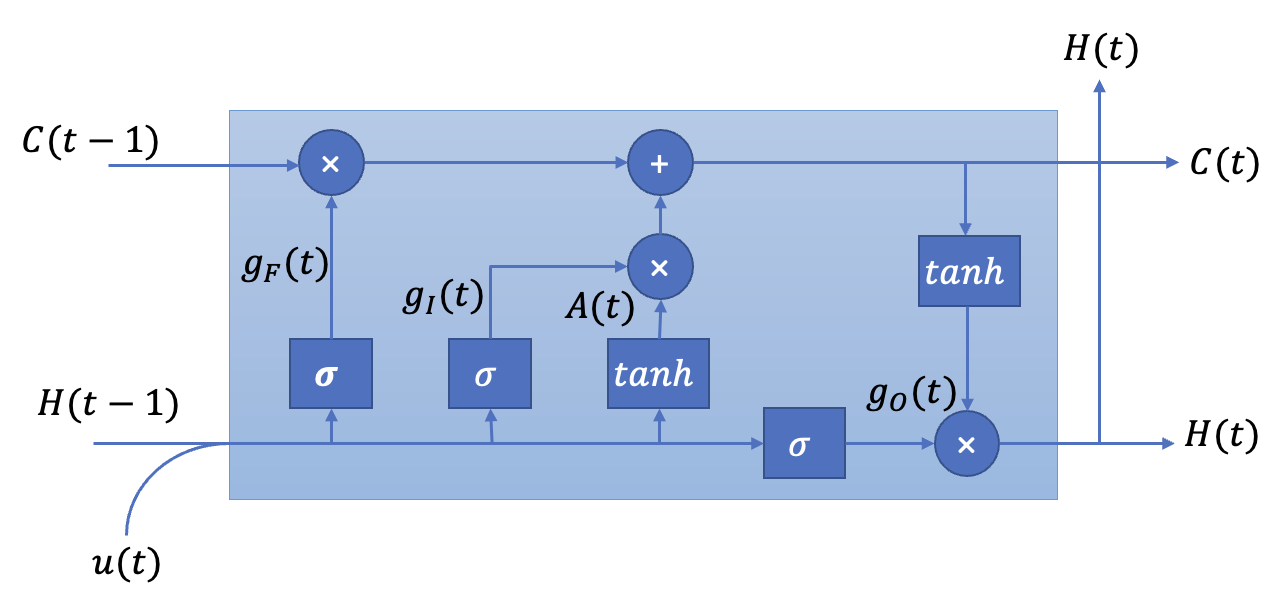}}
\caption{The structure of Long Short Term Memory}
\label{LSTM}
\end{figure}

\begin{equation}
g_F(t) = \sigma (H(t-1)*W_f + u(t)*V_f),
\end{equation}
\begin{equation}
g_I(t) = \sigma (H(t-1)*W_i + u(t)*V_i),
\end{equation}
\begin{equation}
g_o(t) = \sigma (H(t-1)*W_o + u(t)*V_o),
\end{equation}
\begin{equation}
A(t) = tanh (H(t-1)*W_g + u(t)*V_g),
\end{equation}
\begin{equation}
C(t) = g_F(t)*C(t-1) + g_I(t)*A(t),
\end{equation}
\begin{equation}
H(t) = g_o(t)*tanh(C(t)).
\end{equation}

Where $u(t)$ is the input of the network, $H(t)$ is the hidden state in current time step $t$, $\sigma$ points to logistic sigmoid function, $tanh$ is the hyperbolic tangent, $W$ and $V$ are weight matrices that are associated with hidden state and input respectively, and $A(t)$ is cell input activation.

\subsection{Echo State Network (ESN)}
\begin{figure}
    \centering
    \includegraphics[scale=0.5]{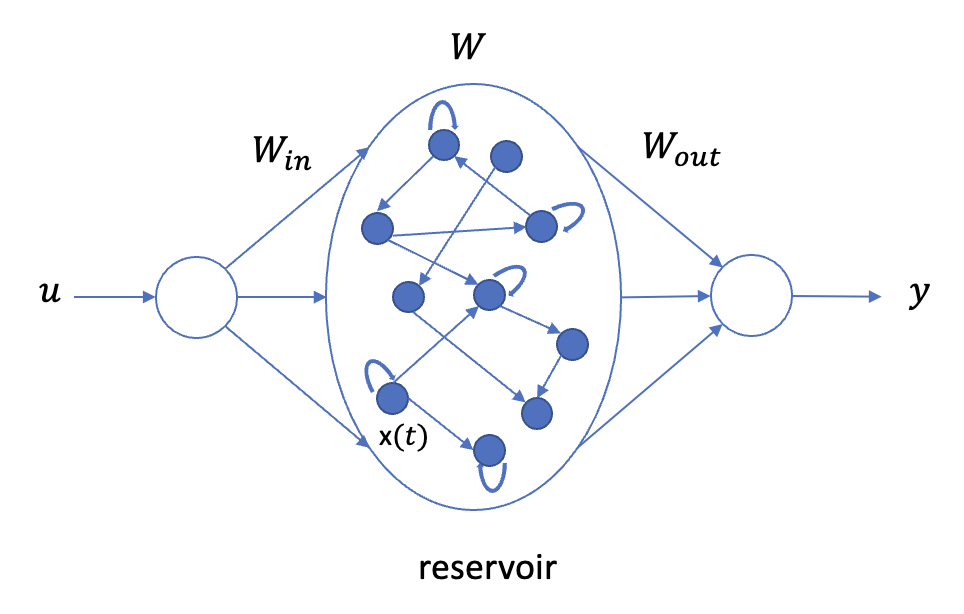}
    \caption{The structure of Echo State Network}
    \label{fig5}
\end{figure}

ESNetwork is a recurrent neural network that uses reservoir computing to make training fast \cite{b3}. This model is shown to be suitable for predicting chaotic time series. Like many other machine learning methods, an ESN model has three layers: one layer as the input layer, one random reservoir, and one layer for output (see \autoref{fig5}). The first step in designing an ESNetwork is to create a large, random reservoir. Connections and weights between neurons in the reservoir are fixed and do not need to be trained. The weights of the input layer are also fixed and chosen randomly. The only trainable parameter of this network is output weights. Each neuron of the reservoir makes a random nonlinear signal. Choosing the weights randomly makes the reservoir like a rich pool of state dynamics that a linear combination can reach the desired output. Hence, these parameters are usually found by linear regression.

In traditional training methods of RNNs(all weights were trainable), most changes caused by gradient algorithms are in the output layer \cite{b2}. So if only output parameters were trained, the problem of gradient vanishing and slow training of RNNs would be solved. 

In this model, inputs connect to reservoir neurons through selected random weights. The following equation updates the neuron's states:

\begin{equation}
x(t) = (1-a)x(t-1) +a tanh(W_{in}u(t)+W x(t-1)),
\label{ESN}
\end{equation}

Where $x$ and $u$ are, respectively, the reservoir state and the network input at time step t, $a$ denotes the leaking rate, $tanh$ is the hyperbolic tangent activation function that is adopted in this paper, $W$ is the matrix of recurrent reservoir weight and $W_{in}$ is the matrix of input weight. The network output is calculated as follows:

\begin{equation}
    y(t) = W_{out}x(t),
    \label{output_ESN}
\end{equation}

Where $W_{out}$ is the matrix of readout weight. There are several known ways to find the optimal output weights $W_[out]$, including: Moore-Penrose pseudo-inverse, regression with Tikhonov regularization\cite{b8}, Bayesian regression\cite{b9}, Levenberg-Marquardt\cite{b10}.

\subsection{Deep Echo State Network}
\begin{figure}[htbp]
\centerline{\includegraphics[scale=0.3]{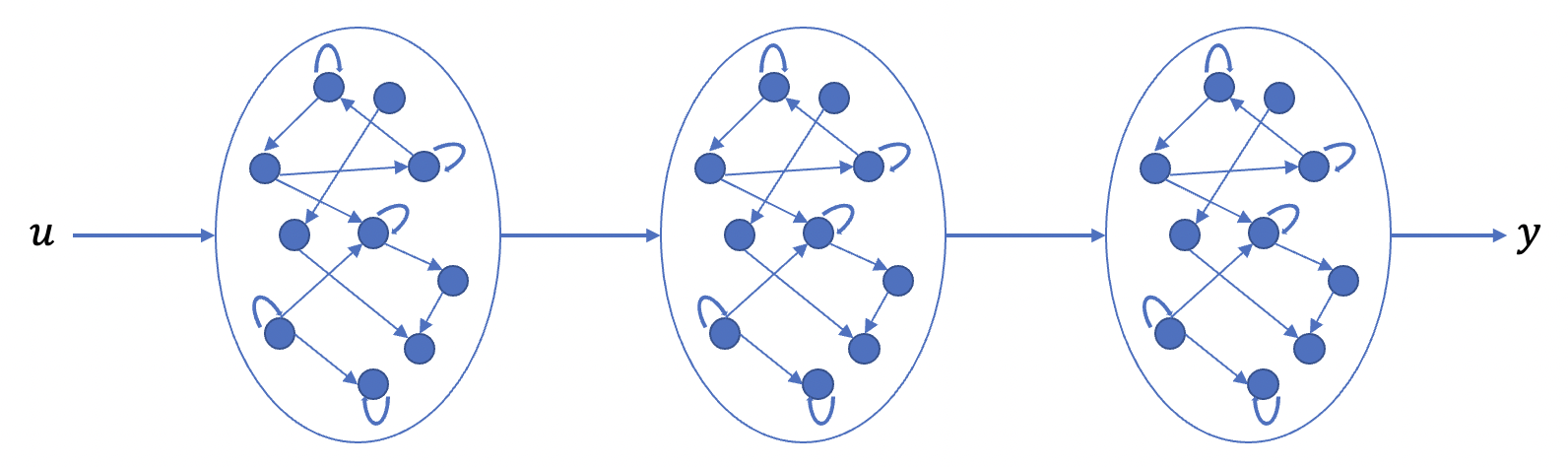}}
\caption{The structure of Deep Echo State Network}
\label{fig4}
\end{figure}

DeepESN can be built in different architectures\cite{b7}. This study focuses on structures in which reservoirs are stacked straight forward. The structure of DeepESN is illustrated in \autoref{fig4}. The input of the first layer is external input, and the inputs of other layers are the previous layer's output. Each reservoir works like an ESN, and each layer's state transition function is as follows
\begin{multline}
  x^{(l)}(t) = (1-a ^{(l)})x^{(l)}(t-1) \\ +a ^{(l)}tanh(W_{in}^{(l)}i^{(l)}(t)
+W^{(l)}x^{(l)}(t-1)),
\label{DeepESN}
\end{multline}

Where the superscript $l$ is used to refer to the network parameter and hyperparameter of layer $l$ and
$i$ refers to the input for the $l$-th layer.
A readout component is used to compute the network output, as told in the ESN section. By taking into account all layers and their states, \autoref{output_ESN} can be rewritten to calculate the output in time step $t$:
\begin{equation}
    y(t) = W_{out}[x^{(1)}(t), x^{(2)}(t), x^{(3)}(t), ..., x^{(N)}(t)]^T,
    \label{DESN out}
\end{equation}
Where N indicates the number of reservoir layers. Methods described in the ESN section are also used here to determine the optimal output weights. For further information about the DeepESN algorithm, refer to \cite{b7}.

\subsection{Radial Basis Function Neural Network (RBF-NN)}
\begin{figure}
    \centering
    \includegraphics[scale=0.5]{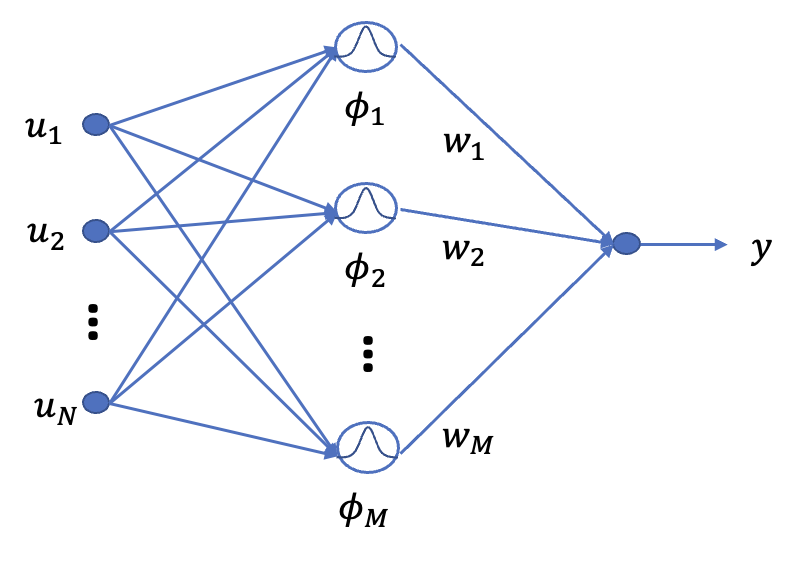}
    \caption{The structure of Radial Basis Function Neural Network}
    \label{RBF}
\end{figure}

RBF ($\phi(x)$) is a function whose value is determined by the Euclidean distance (other types of measures can also be used) to a certain point $x_c$. Typically, it forms as a Gaussian probability density function (\ref{quessi}). The network output is a weighted sum of a set of RBFs. The structure of this network is illustrated in \autoref{RBF}. It is a feed-forward network.

\begin{equation}
\phi(x) = exp(-\frac{||x-x_c||^2}{2\sigma^2}),
    \label{quessi}
\end{equation}

Where $x$ is the input of the RBF network, $x_c$ denotes the center of the basis function, and $\sigma$ is the smoothness of the basis function.

\subsection{Local Linear Model Tree (LoLiMoT)}
LoLiMoT method divides the space of input into some local linear models using an incremental tree. Each division makes the performance of the network better and improve the speed of network computation.

\begin{figure}[htbp]
\centerline{\includegraphics[scale=0.5]{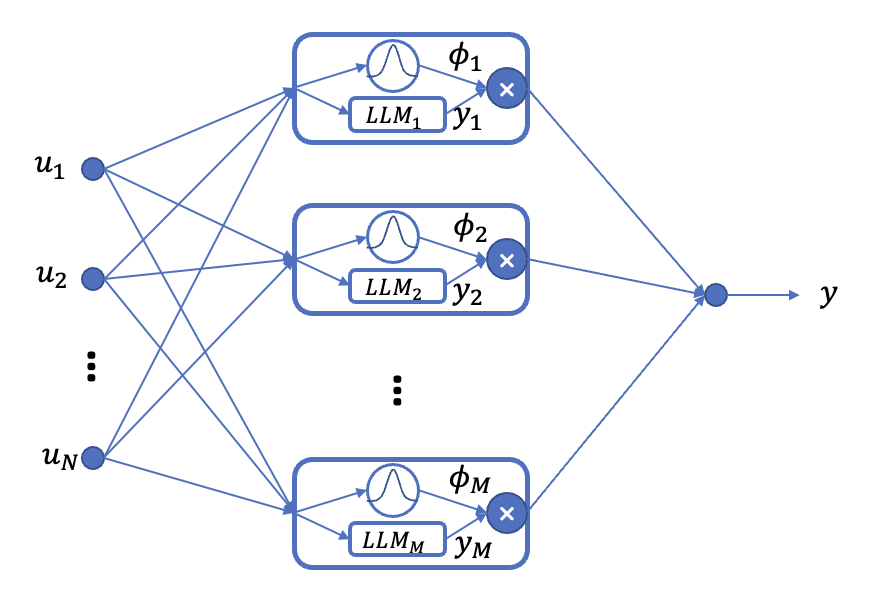}}
\caption{The structure of Local Linear Model Tree (LoLiMoT)}
\label{fig2}
\end{figure}

\begin{figure}[htbp]
\centerline{\includegraphics[scale=0.6]{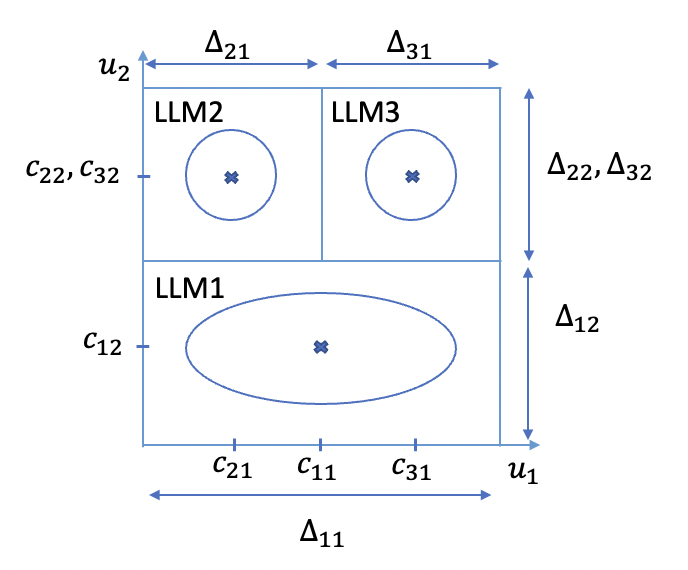}}
\caption{The relationship between input division and its validity function in Local Linear Model Tree (LoLiMoT)}
\label{fig1}
\end{figure}

A network structure diagram can be found in \autoref{fig2}. This network has some neurons in its hidden layer that are composed of a local linear model coupled with a validity function. In most cases, validity functions are selected as normalized Gaussian functions and computed for each division, as shown in \autoref{fig1} \cite{b4}. The output layer of the network has one neuron, and its value is a linear combination of all fuzzy neurons' outputs (local linear models and their validity function).

Training this network is based on the following iterative steps: The first step is to select the worst local linear model based on their loss functions. The Second step is to check all possible divisions of local linear model on input space. In the last step best division choose. It is added as a new neuron in the network. For more information about how it works, refer to \cite{b5}.

\section{Experiment and comparison}
\label{result}
\subsection{Data Analysis}
\begin{figure}[htbp]
\centerline{\includegraphics[scale=0.7]{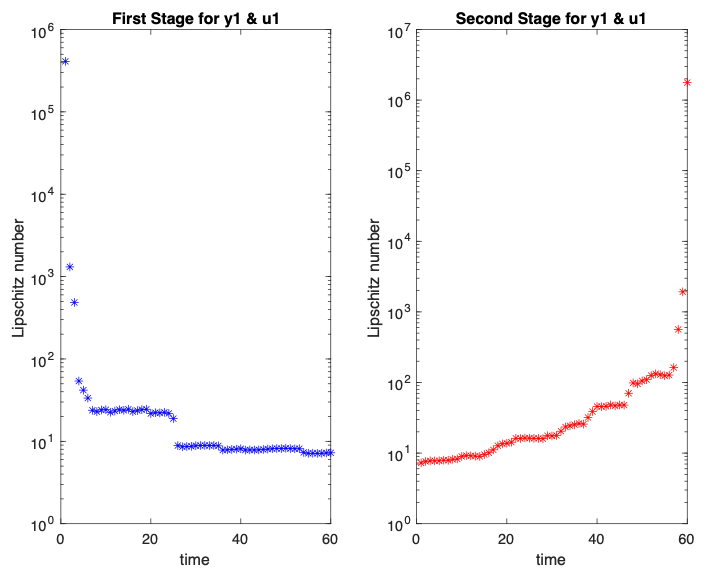}}
\caption{The result of one of LFP signals in input delay estimation}
\label{lipschitz}
\end{figure}

All recorded LFP signals are stationary. It means that the statistical features of the signal do not depend on the time at which the series is observed. Consequently, a model can be found that predicts time series at all times. The first step of every model identification is input delay estimation. A Lipschitz theorem-based method is used for this purpose \cite{b12}. According to this method, a sudden jump in the second stage indicates that output is affected by the last sample that was removed. So, it can be concluded that the first jump estimates the model delay. \autoref{lipschitz} shows the result of estimating the input delay of one of the LFP signals. In the second stage, we can see that the first jump occurs in 2ms due to the sampling time. Therefore, the model delay can be estimated at 2ms. In addition, it is possible to obtain the required dynamics for model identification by using this method. The settling time of the Lipschitz number in the first stage determines the appropriate number of dynamics. In our example, the settling time is approximately 60ms. We should choose at least 60 samples to predict the next samples as plant dynamics.

After finding the required dynamics for model identification, normalize the recorded data of the rat's brain between -1 and 1. In the next step, the dataset is divided into train and test sets, 70\% for the training set and 30\% for the test set. Likewise, all machine learning methods, the training set and the test set are used for training the model and evaluating the model. Ultimately, measurement is required to evaluate the model's performance in time series prediction. In this paper, MAE, which indicates the average absolute error between real and predicted value, is chosen to measure prediction performance.

\subsection{Result}
The present section compares the performances of the methods discussed above in time series prediction. Each system was run with different settings (different parameters and hyperparameters), and the best-achieved performance was used to compare to other methods' performance. For the ESN algorithm, some methods for finding optimal output weights were mentioned before. Obtained weights in Bayesian regression and Moore-Penrose pseudo-inverse methods are near singular, and weights in regularization are not stable. Therefore, Levenberg-Marquardt is used to calculate the output weights.

An analysis of the performance of different methods for predicting time series and identifying the system of the brain of rat7 rewarded by Morphine is provided in \autoref{tab1}. This table shows the results of each network for different time horizons. LoLiMoT has the best performance among the networks mentioned above, as shown in this table. Also, It can be seen that other networks predicted the brain signal at well-enough performance levels. All LFP signals have normal distributions with the Kolmogorov-Smirnov test. This feature helps us identify a model with better performance for the rat's brain among models based on Gaussian functions.

\begin{table}[htbp]
\caption{Comparison of The Results of 
Different Methods in Time Series Prediction}
\begin{center}
\begin{tabular}{|c|c|c|c|}
\hline
\textbf{}&\multicolumn{3}{|c|}{\textbf{Time Horizon(ms)}} \\
\cline{2-4} 
\textbf{Methods} & \textbf{1}& \textbf{5}& \textbf{10} \\
\hline
\textbf{LSTM} & {0.0756}& {0.07971}& {0.08638} \\
\hline
\textbf{ESN} & {0.00442}& {0.05175}& {0.07321} \\
\hline
\textbf{DeepESN} & {0.002507}& {0.04767}& {0.07193} \\
\hline
\textbf{RBF} & {0.0028}& {0.0449}& {0.0743}\\
\hline
\textbf{LoLiMoT} & \textbf{0.0001487}& \textbf{0.0148}& \textbf{0.04943} \\
\hline
\end{tabular}
\label{tab1}
\end{center}
\end{table}

To find a global data-driven model of the rat's brain, we should train a model based on one of the signals that can then be used to predict the other signals. None of the methods, except LoLiMoT, can find a pattern to predict other time series. So, we train a Multi-Input Multi-Output (MIMO) LoLiMoT model with one of the recorded data from one region as the training set. Here, recorded LFP from Nucleus Accumbens region of rat7, which is Morphine rewarded. Furthermore, test the model with other datasets (it does not matter which region is or whether the rat is rewarded or unrewarded). The network outputs are equal to the number of samples that will be predicted, and the number of inputs is equal to the number of plant dynamics. In this work, the network consists of 60 inputs and five outputs. \autoref{tab2} shows the outcome of time series prediction for all signals for 5$ms$ (5 samples). \autoref{tab2} shows that all recorded LFPs have the same dynamical model regardless of whether the rat is rewarded or unrewarded. We can then use a model based on one of the signals to predict the other signals. The obtained model is a global data-driven model based on recorded LFP from rats' Nucleus Accumbens and Hippocampus regions.

\begin{table}[htbp]
\caption{Comparison of The Results of Local Linear Model Tree on
Different Rats}
\begin{center}
\begin{tabular}{|c|c|c|c|c|c|}

\hline
\textbf{}&\textbf{}&\multicolumn{2}{|c|}{\textbf{Pre Test}} &\multicolumn{2}{|c|}{\textbf{Post Test}}\\
\cline{2-6} 
\textbf{} & \textbf{Case}&\textbf{Nac}& \textbf{Hip}& \textbf{Nac} & \textbf{Hip} \\
\hline
\multirow{5}{*}{\rotatebox[origin=c]{90}{Morphine}} & \textbf{Rat1} & {0.01142}& {0.01273}& {0.01315}& {0.01250} \\
\cline{2-6} 
& \textbf{Rat3} & {0.01895}& {0.00460}&  {0.01809}& {0.00762} \\
\cline{2-6}
& \textbf{Rat5} & {0.00751}& {0.01611}&  {0.01182}& {0.00822} \\
\cline{2-6}
& \textbf{Rat6} & {0.01002}& {0.01549}&  {0.01306}& {0.00904} \\
\cline{2-6}
& \textbf{Rat8} & {0.01502}& {0.01152}&  {0.01633}& {0.00977} \\
\Xhline{2\arrayrulewidth}

\multirow{6}{*}{\rotatebox[origin=c]{90}{Natural}} & \textbf{Rat1} & {0.02295}& {0.00845}& {0.01845}& {0.00786} \\
\cline{2-6} 
& \textbf{Rat7} & {0.01532}& {0.01805}&  {0.01524}& {0.01685} \\
\cline{2-6}
& \textbf{Rat8} & {0.01479}& {0.01126}&  {0.01610}& {0.01211} \\
\cline{2-6}
& \textbf{Rat11} & {0.02481}& {0.00630}&  {0.02012}& {0.00383} \\
\cline{2-6}
& \textbf{Rat15} & {0.01232}& {0.01004}&  {0.03123}& {0.01034} \\
\cline{2-6}
& \textbf{Rat16} & {0.00937}& {0.01401}&  {0.01504}& {0.01143} \\
\Xhline{2\arrayrulewidth}

\multirow{7}{*}{\rotatebox[origin=c]{90}{Saline}} & \textbf{Rat9} & {0.01289}& {0.00668}& {0.02274}& {0.00690} \\
\cline{2-6} 
& \textbf{Rat10} & {0.01758}& {0.00973}&  {0.01290}& {0.00650} \\
\cline{2-6}
& \textbf{Rat11} & {0.01598}& {0.00933}&  {0.01297}& {0.00939} \\
\cline{2-6}
& \textbf{Rat12} & {0.01647}& {0.00979}&  {0.01614}& {0.00910} \\
\cline{2-6}
& \textbf{Rat13} & {0.01593}& {0.01528}&  {0.02204}& {0.01129} \\
\cline{2-6}
& \textbf{Rat14} & {0.01279}& {0.01032}&  {0.01906}& {0.01210} \\
\cline{2-6}
& \textbf{Rat15} & {0.00829}& {0.01013}&  {0.01554}& {0.00936} \\
\hline
\end{tabular}
\label{tab2}
\end{center}
\end{table}

\section{Conclusion and discussion}
\label{con}

In conclusion, this paper has introduced a data-driven model capable of predicting rat brain signals. Our model has demonstrated superior performance compared to the current state-of-the-art methods, providing a significant advancement in the field of neural signal prediction. We specifically focused on data collected from the Nucleus Accumbens and Hippocampus areas of rat brains, and our findings have several noteworthy implications. First, one of the most remarkable outcomes of our study is the model's ability to generalize across various scenarios. It successfully predicts brain signals for all rats in the dataset, regardless of whether they were exposed to rewards or remained unrewarded. This robustness indicates that the dynamic features of neurons in these regions remain consistent, regardless of external stimuli, thus highlighting the stability of these neural processes.

Second, we observed that while rewards, such as Morphine or natural rewards, do influence the state of the brain, they may primarily impact the initial states of neurons. Importantly, this finding emphasizes that the core dynamic features of brain signals, as captured by our model, remain unchanged. This insight is crucial for understanding how external stimuli interact with the brain's intrinsic processes.

Last, our research underscores that the shared dynamic features of different brain signals, even in diverse scenarios, do not conflict with the inherent differences in these signals. This revelation challenges the notion that diverse conditions should yield different dynamical neural responses, pointing to a deeper understanding of neural dynamics.

\end{document}